# X-ray fluoresced high-Z (up to Z = 82) K-x-rays produced by $LiNbO_3$ and $LiTaO_3$ pyroelectric crystal electron accelerators

James D. Brownridge[a)]
*Department of Physics, Applied Physics and Astronomy, State University of New York at Binghamton, P.O. Box 6000, Binghamton, New York 13902-6000*

Stephen M. Shafroth[b)]
*Department of Physics and Astronomy, University of North Carolina at Chapel Hill, Chapel Hill, North Carolina 27599-3255*

## Abstract

High-energy bremsstrahlung and K X-rays were used to produce nearly background-free K X-ray spectra of up to 87 keV (Pb) via X-ray fluorescence. The fluorescing radiation was produced by electron accelerators, consisting of heated and cooled cylindrical $LiTaO_3$ and $LiNbO_3$ crystals at mTorr pressures. The newly discovered process of gas amplification whereby the ambient gas pressure is optimized to maximize the electron energy was used to produce energetic electrons which when incident on a W/Bi target gave rise to a radiation field consisting of high-energy bremsstrahlung as well as W and Bi K X-rays. These photons were used to fluoresce Ta and Pb K X-rays.

Although pyroelectric crystals have been known and studied extensively since ancient Greek times[1], interesting properties and innovative applications are still being discovered.[2-25] There is a voluminous literature on electron emission and production phenomena[1-20] and plasma[22-25] formation at the surface of pyroelectric and ferroelectrics crystals. Reference 4, a recent comprehensive review article, is representative but does not suggest that pyroelectric crystals can produce focused, stable energetic electron beams capable of ionizing K electrons in high-Z materials and producing K x-rays. Nor is there a suggestion that these crystals can produce focused ion beams.1[8] Much of the current research attention and industrial use of these crystals is in low energy emission processes, infrared detection, sensitive temperature-change detectors and photonics.[9-11]

There are interesting practical applications of pyroelectric crystal electron accelerators. Since these devices can be relatively inexpensive and safe, they can be used for laboratory teaching[15] of x-ray and energetic electron beam physics. The effects of changes in various parameters such as pressure, temperature, heating and cooling rates etc. can also be studied and controlled by student-written Lab View programs.[16]. These x-rays can also be used to produce radiographs[19]. For example if a sample, such as a broken chicken wing or a small mammal is placed on top of x-ray film and is located a few cm below the x-ray generator a fairly clear radiogram is obtained. With some development it should be possible to fluoresce U and Pu K x-rays for anti-terrorism purposes. Growth processes in plants and tree leaves can be studied by using them as targets and analyzing the electron excited x-rays of elements in the leaves.[21]

The main object of this paper is to show that nearly background-free bremsstrahlung fluoresced K x-rays of elements up to Pb (Z = 82) can be produced by pyroelectric crystal electron accelerators if the process of ”gas amplification", which has been described recently in ref 12 for electrons is taken advantage of. A secondary object is to show that gas amplification occurs on cooling but not on heating for a given crystal by means of the x-ray spectra. The idea is that optimizing the ambient gas pressure can more than double the maximum electron energy. When the -z base of a pyroelectric crystal is exposed to dilute gas during cooling, gas molecules in the vicinity of the crystal are ionized[14] and plasma is formed near the negatively charged surface. The resulting positive ions are attracted to the crystal surface thus reducing its negative charge as well as the electron energy while; the electrons are focused by the plasma and accelerated away from the crystal.[17,18,20] The maximum energy of electrons produced during cooling can be varied by a factor of more than two by varying the gas pressure surrounding the crystal between 0.1 mTorr and about 10 mTorr.[12] On heating a crystal when its surface is positive the gas amplification effect is inoperative and this is demonstrated via the observed x-ray spectra.

When the +z base of a crystal is exposed to dilute gas during cooling positive ions are focused and accelerated away from the crystal while electrons are accelerated to and impinge on the crystal producing x rays and bremsstrahlung at the surface of the crystal[l2]. We used these crystal x-rays and bremsstrahlung to produce relatively background-free fluoresced K x-rays of lower Z elements

Except for the commercially produced pyroelectric crystal x-ray generator by Amptek[19] whose maximum bremsstrahlung energy is about 35 keV, no studies of the behavior of these crystals in dilute gases, for the production of high-Z K x ray production, have previously been reported.

An experimental arrangement which produces the highest energy radiation, is shown schematically in Fig. 1 where a 4 mm dia x 10 mm $LiNbO_3$ crystal with its –z base exposed produces a focused electron beam[18,20] which impinges on a composite target of W/Bi. This gives rise to high-energy bremsstrahlung and W/Bi K x-rays, which were used to fluoresce K X-rays of Ta and Pb.

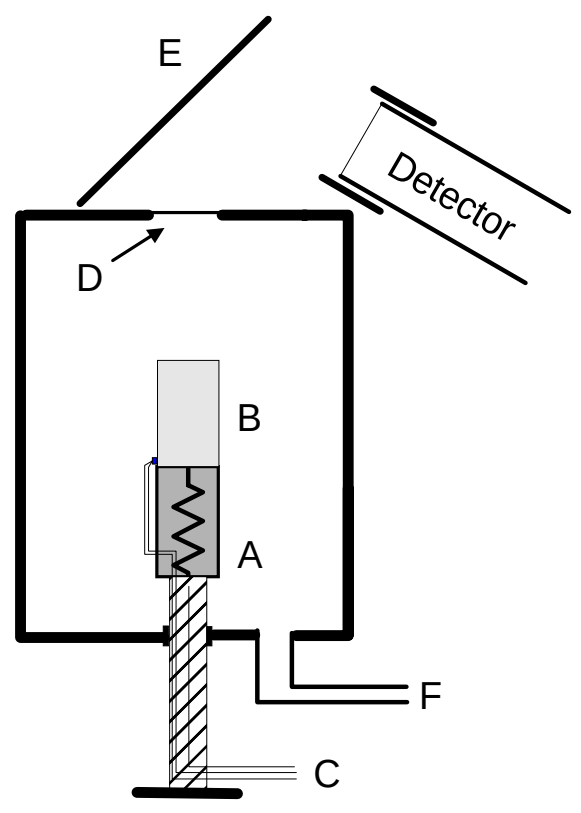

FIG. 1. A thin target D of W/Bi is epoxied to a 2.54 $10^{-3}$ cm Al foil is irradiated by energetic electrons from crystal B on cooling and gives rise to K X-rays and bremsstrahlung. This radiation fluoresces a target at E. The Si/Li detector outside the chamber records the resulting X-ray spectra.

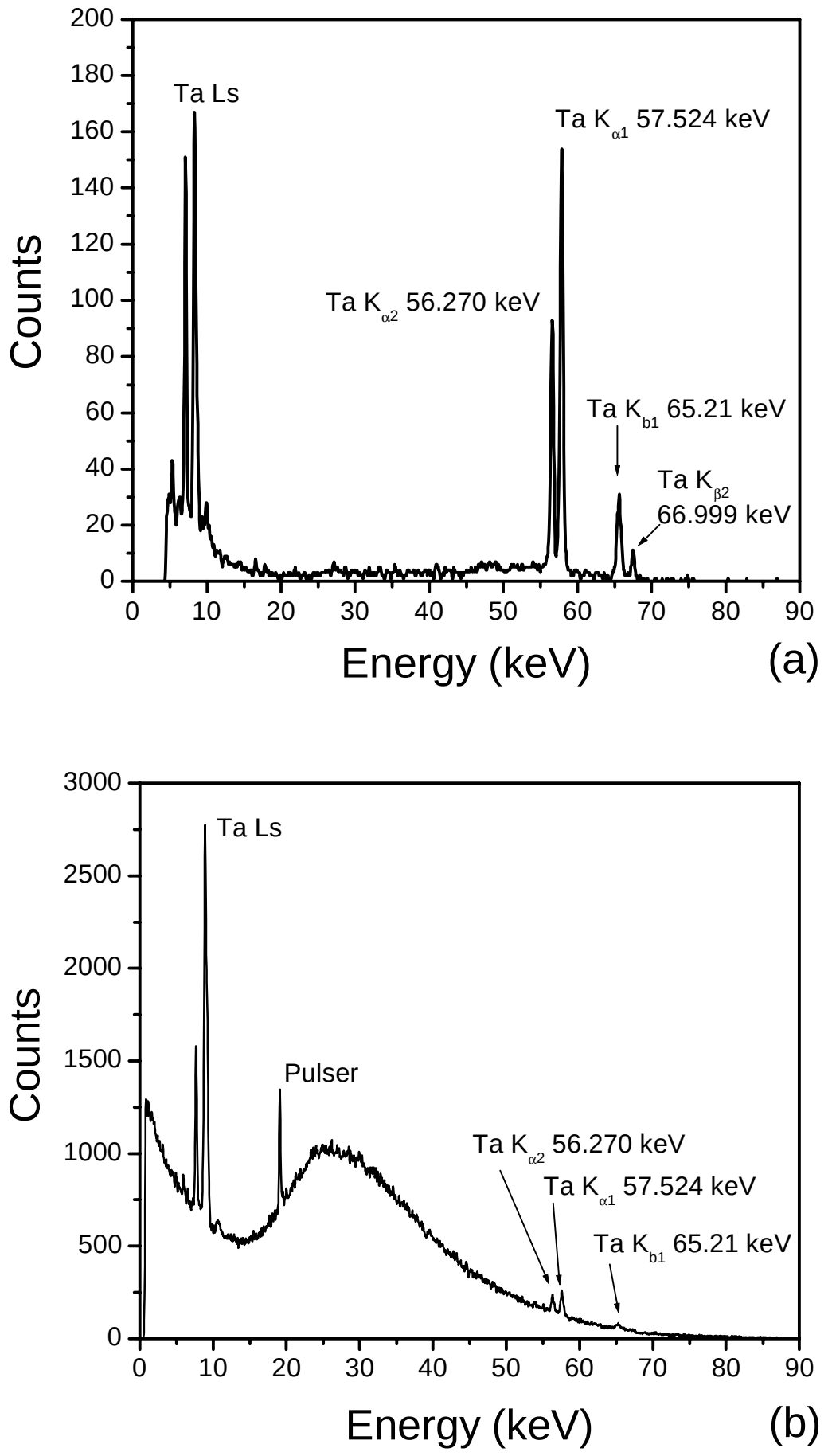

FIG. 2. (a) Bremsstrahlung- fluoresced spectrum of Ta when the geometry is as shown in Fig. 1 and the crystal is $LiNbO_3$. Note the background-free nature of the spectrum. Only one thermal cycles was used and the x-ray detector was only activated when the incident electron energy exceeded the Ta K edge, 67.4 keV. (b) Shows Ta K X-rays superimposed on a bremsstrahlung background when the Ta is irradiated directly by electrons rather than by photons.

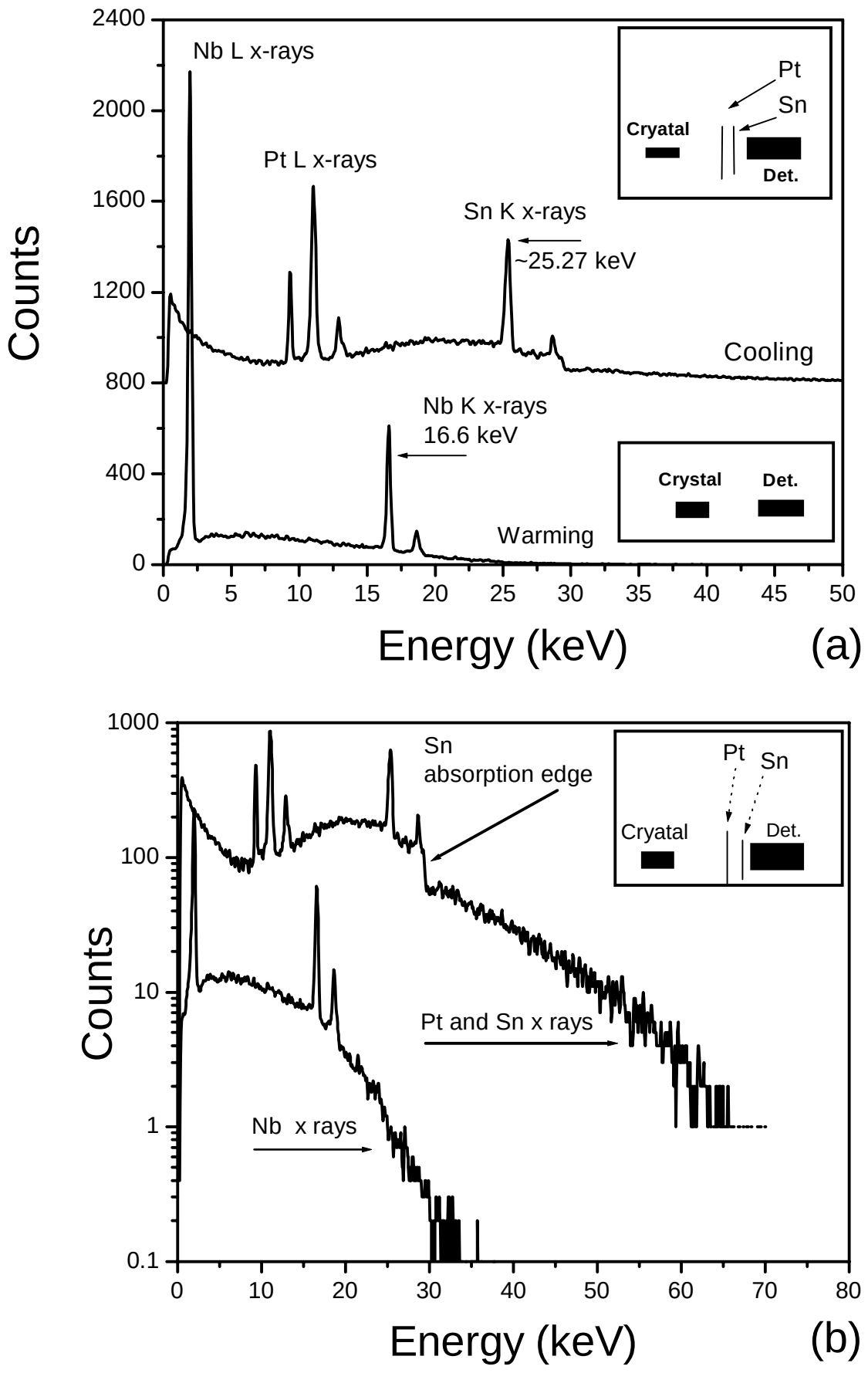

FIG. 3. (a) Bremsstrahlung fluoresced Sn K x-rays superimposed on attenuated Bremsstrahlung spectrum arising from electron bombardment of Pt. The crystal was $LiNbO_3$ (5mm long) with its -z base exposed. Note Pt electron excited L x-rays. The geometry is collinear as shown above in the inset. The logarithmic Pt/Sn spectrum in 3(b) is multiplied by an arbitrary number for display purposes. The gas amplification effect is demonstrated since the “cooling” bremsstrahlung maximum energy end point is twice as high as the “heating” end point energy.

The bremsstrahlung and W and Bi K X-ray fluoresced spectrum of a metal foil of Ta is shown in Fig. 2(a) The K x-ray spectrum is very free of background because the cross section for K-shell ionization by photons whose energy is just above the Ta K-shell ionization energy is much greater than the cross section for Compton and elastic scattering of the lower and higher energy bremsstrahlung by the Ta target.

Fig. (2b) shows the spectrum obtained when electrons rather than photons directly bombard the Ta target. The peak-to-background ratio is much smaller. The Ta L x-ray spectrum is very prominent too but the peak-to-background ratio is not as good as for the K x-rays. The peak to background ratio for this spectrum is more than a factor of two better than the Ta K X-ray spectrum obtained with 5.6 MeV gamma rays at the Duke Free Electron Laser (FEL).[26]

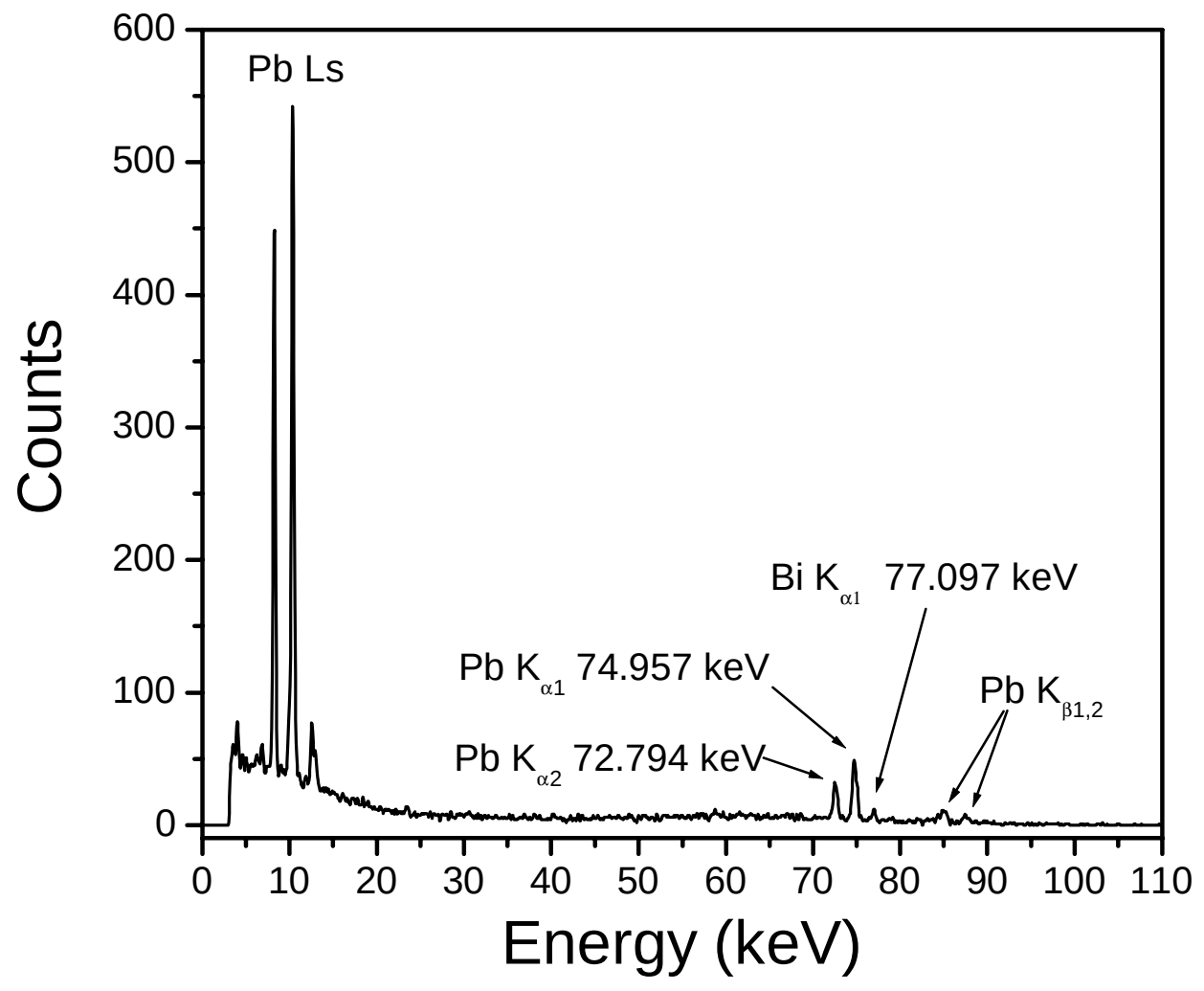

FIG. 4. X-ray spectrum of bremsstrahlung fluoresced Pb combined with Bi K x-rays from the target. The geometry was that of Fig. 1 The 10 mm $LiNbO_3$ crystal provided higher energy electrons than the 5 mm $LiTaO_3$ crystal. This spectrum shows that the electron energy was > 90.5 keV, the Bi K-shell ionization energy.

Bremsstrahlung fluoresced Sn K x-rays superimposed on attenuated Bremsstrahlung spectrum arising from electron bombardment of Pt is shown in a linear display in Fig. 3(a). The geometry is collinear as shown above in the inset. The log display of the Pt/Sn spectrum in Fig. 3(b) is multiplied by an arbitrary number for display purposes. The electron-excited Pt L x-rays, which are attenuated by the Sn are prominent. The crystal was $LiNbO_3$ (5mm long) with its -z base exposed and the Nb K x-ray spectrum, which was produced during heating to 160$^o$C, is shown as well as the asymmetry of gas amplification. The Pt/Sn target was removed during heating for this run to permit observation of Nb L x-rays. The Pt/Sn target removal had minimum effect on the end point energy. We have observed gas amplification only on cooling when the –z base is exposed, the bremsstrahlung end point energy in this case is about 65 keV while in the heating mode, at the same pressure, the bremsstrahlung end point energy is about 33 keV. While the end point energy will change with pressure on cooling, during heating the end point energy is virtually unchanged. This suggests that there is no plasma energy-enhancing process occurring when the crystal surface is positive, while when the crystal surface is negative a plasma energy-enhancing effect occurs. When the crystal's surface is negative and the crystal is cooling i.e., the –z base is exposed, the plasma at the surface of the crystal may be displaced in several ways. A gradual or sudden change in the gas pressure causes the plasma to be expelled from the crystal. This is illustrated in reference 12, Figs. 2 and 3, reference 14 in Figs. 6, 7 and 9 and reference 20. The plasma may also be separated from the crystal by a sudden violent movement of the crystal from a stationary position or by increasing the angular velocity when the crystal is rotating in dilute gas.

Figure 4 shows the x-ray spectrum of bremsstrahlung fluoresced Pb combined with leak-through Bi K x-rays from the W/Bi target. The count rate was very low here so two thermal cycles were required and the detector could view a little bit of the target, the source of the Bi. Nevertheless this proves that the electron energy was > 90.5 keV, the Bi K-shell ionization energy.

In conclusion we show how heated and cooled cylindrical pyroelectric crystals such as $LiNbO_3$ and $LiTaO_3$ under the right conditions can be used to produce relatively background-free characteristic K x-ray spectra for high-Z elements. We compare these spectra with electron-excited spectra, where the background is much larger, and we show how the gas amplification process can be observed via x-rays when the crystal surface is negative whereas no electron-energy enhancement occurs when the crystal surface is positive.

We are most grateful to our colleagues, Sol Raboy, Tom Clegg and Eugen Merzbacher for insightful discussions, and continued encouragement.

[a)]e-mail jdbjdb@binghamton.edu
[b)]e-mail shafroth@physics.unc.edu